\documentclass[conference,a4paper]{IEEEtran}
\usepackage{xcolor}
\usepackage{balance}
\usepackage{cite}
\usepackage{amsmath,amssymb,amsfonts}
\usepackage{graphicx}
\usepackage{textcomp}
\usepackage{acronym}
\usepackage{xcolor}
\usepackage{tikz} 
\usepackage[utf8]{inputenc}
\usepackage{pgfplots} 
\usepackage{pgfgantt}
\usepackage{pdflscape}
\usepackage{changes}
\usepackage{comment}
\usepackage{subfigure}
\usepackage{bm}
\usepackage{bbm}
\usepackage[noend]{algpseudocode}
\usepackage{mathtools,algpseudocode,algorithm,MnSymbol}

\usepackage{geometry}
\usepackage{pgfplots}
  \pgfplotsset{compat=newest}
  \usetikzlibrary{plotmarks}
  \usetikzlibrary{arrows.meta}
  \usepgfplotslibrary{patchplots}
  \usepackage{grffile}
  \usepackage{amsmath}

\pgfplotsset{compat=newest} 
\pgfplotsset{plot coordinates/math parser=false} 

\acrodef{MMSE}{Minimum Mean Squared Error}

\acrodef{MSE}{mean square error}

\acrodef{PSD}{power spectral density}

\acrodef{RMSE}{root mean squared error}
\acrodef{SLR}{statistical linear regression}

\acrodef{ML}[ML]{maximum likelihood}
\acrodef{DBSCAN}[DBSCAN]{density-based spatial clustering of applications with noise}
\acrodef{UE}[UE]{user equipment}
\acrodef{BS}[BS]{base station}
\acrodef{VA}[VA]{virtual anchor}
\acrodef{SP}[SP]{scattering  point}
\acrodef{IP}[IP]{incidence point}
\acrodef{fov}[FoV]{field-of-view}   
\acrodef{LoS}[LoS]{line-of-sight}
\acrodef{NLoS}[NLoS]{non-line-of-sight}
\acrodef{PMBM}[PMBM]{Poisson  multi-Bernoulli  mixture}
\acrodef{PMB(M)}[PMB(M)]{Poisson  multi-Bernoulli  (mixture)}
\acrodef{PMB}[PMB]{Poisson  multi-Bernoulli}
\acrodef{RFS}[RFS]{random finite set}
\acrodef{PPP}[PPP]{Poisson point process}
\acrodef{MBM}[MBM]{multi-Bernoulli  mixture}
\acrodef{MB}[MB]{multi-Bernoulli}

\acrodef{ekf}[EKF]{extended Kalman filter}
\acrodef{PDF}[PDF]{probability density function}

\acrodef{ckf}[CKF]{cubature Kalman filter}
\acrodef{rbp}[RBP]{Rao-Blackwellized particle}
\acrodef{gospa}[GOSPA]{generalized optimal subpattern assignment}
\acrodef{SLAM}[SLAM]{simultaneous localization and mapping}

\acrodef{TOA}[TOA]{time of arrival}
\acrodef{AOA}[AOA]{angles of arrival}
\acrodef{AOD}[AOD]{angles of departure}

\acrodef{IF}[IF]{information filter}
\acrodef{EIF}[EIF]{extended information filter}
\acrodef{kf}[KF]{Kalman filter}
\acrodef{PMF}[PMF]{probability mass function}
\acrodef{MAP}[MAP]{maximum a posteriori estimation}

\acrodef{DA}[DA]{data association}
\acrodef{OFDM}[OFDM]{Orthogonal Frequency Division Multiplexing}

\acrodef{PCRB}[PCRB]{posterior Cram{\'e}r-Rao bound}

\acrodef{EM}[EM]{expectation-maximization}
\acrodef{FOV}[FOV]{field of view}
\acrodef{RB}[RB]{Rao-Blackwellized }

\acrodef{ISAC}[ISAC]{integrated sensing and communication}

\acrodef{EK}[EK]{extended Kalman}
\acrodef{EKF}[EKF]{extended Kalman filter}

\acrodef{LMB}[LMB]{labeled multi-Bernoulli}
\acrodef{GLMB}[$\delta$-GLMB]{$\delta$-generalized labeled multi-Bernoulli}
\acrodef{PHD}[PHD]{probability hypothesis density}

\acrodef{OID}[OID]{optimal importance density}

\acrodef{MTT}[MTT]{multi-target tracking}

\acrodef{MCMC}[MCMC]{Markov chain Monte Carlo}

\acrodef{MH}[MH]{Metropolis-Hastings}

\acrodef{NMI}[NMI]{normalized mutual information}

\acrodef{CRB}[CRB]{Cram{\'e}r-Rao bound}

\acrodef{PIM}[PIM]{posterior information matrix}

\acrodef{OID}[OID]{optimal importance density}

\acrodef{MAE}[MAE]{mean absoluate error} 

\acrodef{FIM}[FIM]{Fisher information matrix}
\acrodef{PIM}[PIM]{posterior information matrix}
\acrodef{PEB}[PEB]{position error bound}
\acrodef{LEB}[LEB]{landmark error bound}
\acrodef{HEB}[HEB]{heading error bound}
\acrodef{CEB}[CEB]{clock bias error bound}

\acrodef{AoA}[AoA]{angles of arrival}
\acrodef{AoD}[AoD]{angles of departure}

\acrodef{URA}[URA]{uniform rectangular array}

\acrodef{RTT}{round-trip-time}

\acrodef{LS}{least-squares}

\acrodef{CDF}{cumulative distribution function}

\newcommand{\herm}{^\mathrm{H}}

\newcommand{\vect}{\mathrm{vec}}

\makeatletter
\newcommand*\rel@kern[1]{\kern#1\dimexpr\macc@kerna}
\newcommand*\widebar[1]{%
  \begingroup
  \def\mathaccent##1##2{%
    \rel@kern{0.8}%
    \overline{\rel@kern{-0.8}\macc@nucleus\rel@kern{0.2}}%
    \rel@kern{-0.2}%
  }%
  \macc@depth\@ne
  \let\math@bgroup\@empty \let\math@egroup\macc@set@skewchar
  \mathsurround\z@ \frozen@everymath{\mathgroup\macc@group\relax}%
  \macc@set@skewchar\relax
  \let\mathaccentV\macc@nested@a
  \macc@nested@a\relax111{#1}%
  \endgroup
}
\makeatother

\newcommand{\JJ}{\boldsymbol{J}}

\newcommand{\AAb}{\boldsymbol{B}}
\newcommand{\yym}{\bm{\mathcal{Y}}}
\newcommand{\nnm}{\bm{\mathcal{N}}}
\newcommand{\dd}{\boldsymbol{d}}

\newcommand{\aab}{\boldsymbol{a}}

\newcommand{\aabx}{\aab_{\rm{x}}}
\newcommand{\aabz}{\aab_{\rm{z}}}

\newcommand{\deltaf}{\Delta_f}

\newcommand{\normsmall}[1]{\big\lVert#1\big\rVert}

\newcommand{\aabxw}{\widetilde{\aab}_{\rm{x}}}
\newcommand{\aabzw}{\widetilde{\aab}_{\rm{z}}}

\newcommand{\yymt}{\widetilde{\yym}}

\newcommand{\aabxi}{{\aab}_{{\rm{x}}}^{t}}
\newcommand{\aabzi}{{\aab}_{{\rm{z}}}^{t}}
\newcommand{\ddi}{{\dd}^{t}}

\newcommand{\Mx}{N_{\rm{x}}}
\newcommand{\Mz}{N_{\rm{z}}}
\newcommand{\dz}{d_{\rm{z}}}
\newcommand{\dx}{d_{\rm{x}}}

\newcommand{\rmel}{{{\rm{el}}}}
\newcommand{\rmaz}{{{\rm{az}}}}

\newcommand{\thetaaz}{\theta_{\rmaz}}
\newcommand{\thetael}{\theta_{\rmel}}

\makeatletter
\newcommand{\gettikzxy}[3]{%
  \tikz@scan@one@point\pgfutil@firstofone#1\relax
  \edef#2{\the\pgf@x}%
  \edef#3{\the\pgf@y}%
}

\newcommand{\complexset}[2]{ \mathbb{C}^{#1 \times #2}  }

\newcommand{\complexsett}[3]{ \mathbb{C}^{#1 \times #2 \times #3}  }

\newcommand{\abs}[1]{\big\lvert #1 \big\rvert}

\begin{document}

\bibliographystyle{IEEEtran}
\bstctlcite{IEEEexample:BSTcontrol}

\title{Pilot-Based End-to-End Radio Positioning and Mapping for ISAC: Beyond Point-Based Landmarks}

\author{\IEEEauthorblockN{
Yu Ge\IEEEauthorrefmark{1}, Musa Furkan Keskin\IEEEauthorrefmark{1}, Hui Chen\IEEEauthorrefmark{1}, Ossi Kaltiokallio\IEEEauthorrefmark{2}, Mengting Li\IEEEauthorrefmark{1}\IEEEauthorrefmark{3}, \\Mikko Valkama\IEEEauthorrefmark{2}, 
   Christos Masouros\IEEEauthorrefmark{4},  
Henk Wymeersch\IEEEauthorrefmark{1}     
}                                     
\IEEEauthorblockA{\IEEEauthorrefmark{1}
Chalmers University of Technology, Sweden,   \IEEEauthorrefmark{2}
Tampere University, Finland,}
\IEEEauthorblockA{\IEEEauthorrefmark{3}
Aalborg University, Denmark, \IEEEauthorrefmark{4}
University College London, The United Kingdom}
}

\maketitle

\begin{abstract}
\Acl{ISAC} enables simultaneous communication and sensing tasks, including precise radio positioning and mapping, essential for future 6G networks. Current methods typically model environmental landmarks as isolated incidence points or small reflection areas, lacking detailed attributes essential for advanced environmental interpretation. This paper addresses these limitations by developing an end-to-end cooperative uplink framework involving multiple base stations and users. Our method uniquely estimates extended landmark objects and incorporates obstruction-based outlier removal to mitigate multi-bounce signal effects. Validation using realistic ray-tracing data demonstrates substantial improvements in the richness of the estimated environmental map.

\end{abstract}

\vskip0.5\baselineskip
\begin{IEEEkeywords}
 ISAC, 5G/6G, positioning and mapping, extended object, outlier removal, raytracing.
\end{IEEEkeywords}

\section{Introduction}
\Ac{ISAC}, which enables the radio system to simultaneously provide communication and sensing abilities, has gained significant attention in the evolution of 5G mobile radio systems and has become a pivotal pillar of 6G networks \cite{An_ISAC_Survey2022,liu2022integrated,gonzalez2024integrated}. Sensing using radio signals is at the heart of \ac{ISAC}, which  encompasses the estimation of the states of both connected devices and passive objects in the propagation environment, also termed as radio \emph{positioning} and \emph{mapping}, respectively \cite{chaccour2022seven,ge2023mmwave}. 

Intense research activities in radio positioning and mapping have been taking place, due to its wide range of foreseen applications \cite{Amjad2023,ge2023mmwave}. 
High time and angular resolutions are enabled by the large available bandwidth and antenna arrays in 5G and 6G systems. By exploiting geometric information from the propagation paths, including both \ac{LoS} and \ac{NLoS} paths \cite{leitinger2019belief,ge2022computationally}, the \ac{UE} trajectory and the map of reflecting or scattering objects in the environment can be jointly estimated. Real-world experiments in \cite{ge2024experimental,rastorgueva2024millimeter} demonstrate the feasibility of radio positioning and mapping, which rely solely on transmissions between a single \ac{BS} and a single \ac{UE}. 
However, all these works either model the landmarks in the environment as single points and only estimate the positions of points, or recover the \acp{IP} of the transmitted signals with the landmarks. However, such a way of modeling cannot convey richer attributes of landmarks such as object size, shape, types, and others, all of which are crucial to higher‑level semantic understanding and digital radio twins \cite{khan2022digital}.

Radio positioning and mapping considering an extended object model, which contains richer attributes of landmarks than only position and considers multiple measurements from the same landmark, has also been explored, for example, in \cite{ge20205GSLAM,karttunen2025towards,zhai2025multipath}. However, \cite{ge20205GSLAM,karttunen2025towards} support simultaneous UE localization and landmark mapping, and can distinguish different types of landmarks, but other features of the landmarks, such as size or shape, are not considered. The approach in \cite{karttunen2025towards} associates a small reflection/scattering area to each estimated IP, but landmarks are still represented as collections of mapped IPs. The sizes and shapes of landmarks are considered in \cite{zhai2025multipath} by incorporating a probabilistic framework; however, the problem is not solved in an end-to-end manner. In addition, this approach assumes only single-bounce paths and considers a single UE without any cooperation.

\begin{figure}
    \centering
\includegraphics[width=0.88\columnwidth]{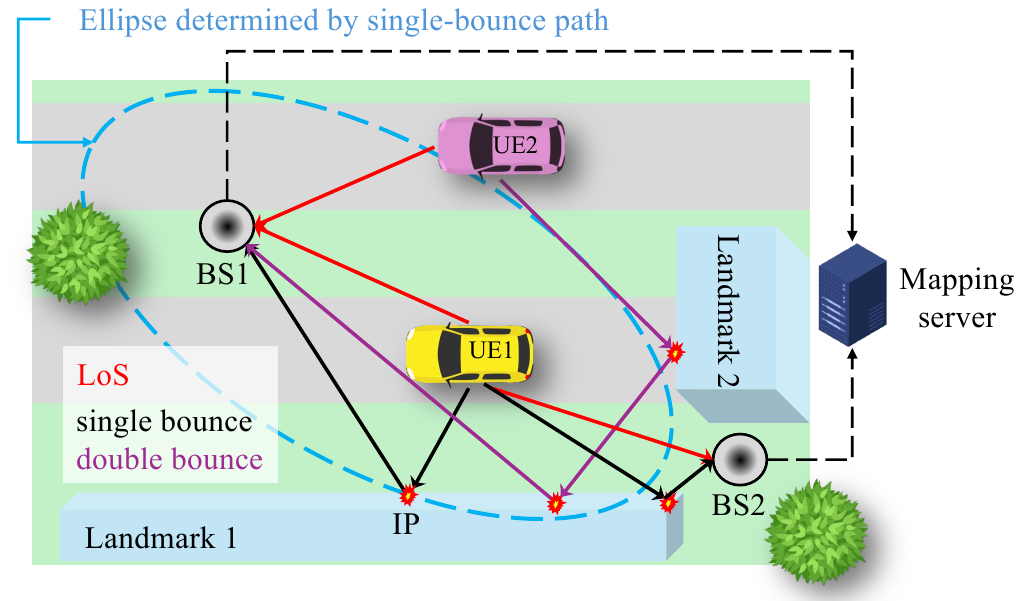}
    \vspace{-3mm} 
    \caption{An example of the considered scenario with 2 BSs and 2 UEs. It also provides an illustrative demonstration of the closed-form solution for estimating the IP of a single-bounce path.}
    \label{fig:ellipse}
\vspace{-5mm} 
\end{figure}

In this paper, we study the radio positioning and mapping problem in the multi-\ac{UE}-multi-\ac{BS} uplink system, as illustrated in Fig~\ref{fig:ellipse}, where each BS localizes its connected UEs on its own and all BSs cooperate with each other by sharing its own estimated \acp{IP} to a fusion center to create a more refined map. The problem is solved in an end-to-end manner, which includes a channel modeling, a high-resolution channel estimator, a low-complexity positioning and mapping algorithm for each BS, and a post-processing mapping refinement process for the fused map. The main contributions of this paper are as follows: \textit{(i)} We propose a novel method that can estimate the extended objects, not only separate points, in the field of radio mapping; \textit{(ii)} We develop an outlier removal method which is based on obstruction detection to remove the artificial landmarks caused by multi-bounce paths; \textit{(iii)} We validate the effectiveness of the end-to-end framework using realistic ray-tracing data, demonstrating the ability of the proposed post-processing algorithm to create more subtle map.

\emph{Notations:}
Scalars (e.g., $x$) are denoted in italic, vectors (e.g., $\boldsymbol{x}$) in bold lower-case letters, matrices (e.g., $\boldsymbol{X}$) in bold capital letters, sets  (e.g., $\mathcal{X}$) in calligraphic. Transpose is denoted by $(\cdot)^{\top}$, the Hermitian transpose is denoted by $(\cdot)^{\mathsf{H}}$, and the outer product is denoted by $\circ$.

\vspace{-1mm}

\section{Signal model}
In this paper, we consider a multi-\ac{BS} and multi-\ac{UE} scenario, where each \ac{BS} performs bistatic sensing using uplink signals to localize \acp{UE} and map the surrounding environment. After that, multiple \acp{BS} share sensing results to a fusion center to create more refined results. In this section, the state models and the received signal model are introduced. 

\vspace{-1mm}

\subsection{State Model}
In the considered environment, there are multiple \acp{UE} moving over time, and each UE is equipped with a single antenna. The state of the $i$-th UE at the time step $k$ is described by the 3D position of its antenna in the global reference coordinate system as $\boldsymbol{x}^{i}_{\text{UE},k}=[x^{i}_{\text{UE},k},y^{i}_{\text{UE},k},z^{i}_{\text{UE},k}]^{\top}$.  Neither the total number of \acp{UE} nor the specific UE states are known to the  \acp{BS}. Unlike UEs, all \acp{BS} are fixed over time, and each \ac{BS} is equipped with a \ac{URA}. The state of the $p$-th BS contains a 3D location $\boldsymbol{x}_{\text{BS}}^{p}=[x^{p}_\text{BS},y^{p}_\text{BS},z^{p}_\text{BS}]^{\top}$, which denotes the location of the center of its \ac{URA} in the global reference coordinate system, and a 3D orientation $\boldsymbol{\psi}^{p}_\text{BS}=[\varepsilon^{p}_\text{BS},\nu^{p}_\text{BS},\gamma^{p}_\text{BS}]^{\top}$, which is ordered as roll, pitch, and yaw, respectively, and denotes the Euler angles of the orientation of the \ac{URA} with respect to the global reference coordinate system. The total number of \acp{BS} and specific \ac{BS} states are known. Each BS has a limited coverage, and only provide positioning services for UEs in its \ac{fov}. Apart from the BSs, there are other unknown landmarks in the considered environment. The state of the $m$-th landmark can be parameterized as a convex shape, denoted as $\mathcal{H}^{m}$, which contains a list of points that describe the convex hull of the landmark.

\subsection{Received Signal Model} \label{Measurement model}
In this paper, we follow the \ac{RTT} protocol, where a UE responds directly to the corresponding BS when it receives downlink signals from that BS \cite{ge2024v2x}. Suppose that at time step $k$, the $p$-th \ac{BS} receives uplink singals from the $i$-th \ac{UE}, and the signal can arrive via a  \ac{LoS} path and/or via \ac{NLoS} paths resulting from reflections, diffractions, or diffusions off landmarks. Each landmark can generate multiple \ac{NLoS} paths, and the signals can interact with several landmarks before reaching the BS. Considering a short transmission interval (neglecting Doppler effects) and \ac{OFDM} transmissions with $N_{\text{OFDM}}$ symbols and $S$ subcarriers, the received signal at the $p$-th BS from the $i$-th \ac{UE} for the $g$-th \ac{OFDM} symbol at the $\kappa$-th subcarrier and time step $k$ can be expressed as \cite{heath2016overview}\footnote{For notational simplicity, we omit the indices of the BS and UE. Unless otherwise stated, all descriptions in the remainder of this section refer specifically to the transmission between the $p$-th BS and the $i$-th \ac{UE}.}
\begin{align}
    \boldsymbol{y}_{\kappa,g,k}&=  \sum _{m=0}^{M_k}\sum _{l=1}^{L_k^{m}}\rho^{m,l}_{k}\boldsymbol{a}(\boldsymbol{\theta}_{k}^{m,l})e^{-\jmath 2\pi \kappa \Delta_f \tau_{k}^{m,l}} s_{\kappa,g} +  \boldsymbol{\omega}_{\kappa,g,k},\label{eq:signalModel} 
\end{align}
where $\boldsymbol{y}_{\kappa,g,k}$ is the received signal across the BS antenna array at the subcarrier $\kappa$, $s_{\kappa,g}$ is the precoded pilot signal at the subcarrier $\kappa$ at the UE side, $\boldsymbol{\omega}_{\kappa,g,k}$ is the additive white Gaussian noise,  $\boldsymbol{a}(\cdot)$ is the steering vector of the BS antenna arrays, and $\Delta f$ is the subcarrier spacing. There are $M_{k}+1$ visible landmarks at time step $k$, with $m=0$ representing the BS, and $L_{k}^{m}$ is the number of paths from each landmark. Each path can be characterized by a complex gain $\rho_{k}^{m,l}$, a \ac{TOA} $\tau_{k}^{m,l}$, and an \ac{AOA} pair $\boldsymbol{\theta}_{k}^{m,l}=[{\theta}_{\text{az},k}^{m,l},{\theta}_{\text{el},k}^{m,l}]^{\top}$ in azimuth and elevation. 
The \ac{AOA} pair $\boldsymbol{\theta}_{k}^{m,l}$ is determined by the direction of the signals that arrive at the receiver for the $(m,l)$-th path, and the \ac{TOA} $\tau_{k}^{m,l}$ is determined by the propagation distance of the $(m,l)$-th path (the clock offset between the transmitter and the receiver is assumed to be cancelled out through the RTT protocol).

\section{End-to-end Process}
To address the positioning and mapping problem, the system exploits the received signals described in \eqref{eq:signalModel} to localize the UEs and map the surrounding environment. This section introduces two key components: the channel parameter estimator, which converts received signals into geometric information, and the snapshot positioning and mapping algorithm, which estimates the positions of UEs and the locations of \acp{IP}.
\subsection{Channel Parameter Estimation} \label{channel_estimator}
\subsubsection{3-D Tensor Observations}
We present an ESPRIT-based high-resolution channel estimator to extract geometric path parameters. In this model, the constant modulus pilots $s_{\kappa,g}$ are eliminated through conjugate multiplication.  
The signal expression in \eqref{eq:signalModel} is reformulated into a three-dimensional tensor $\yym_k \in \complexsett{\Mz}{\Mx}{S}$, given by
\begin{align} \label{eq_yy_3d}
     \yym_k = \sum _{t=1}^{T_{k}}\rho^{t}_{k} \aabz(\boldsymbol{\theta}_{k}^{t}) \circ \aabx(\boldsymbol{\theta}_{k}^{t}) \circ \dd(\tau_{k}^{t}) + \nnm_k \,,
\end{align}
where $T_{k}=\sum _{m=0}^{M_k}L_k^{m}$ denotes the number of all paths at time step $k$, and $\Mx$ and $\Mz$ are the numbers of elements along the horizontal and vertical axes of the URA, respectively, and the integration gains over multiple symbols are encapsulated within $\rho^{t}_{k}$. The spatial steering vectors in the horizontal and vertical domains are defined, respectively, as $\aabx(\boldsymbol{\theta}) \in \complexset{\Mx}{1}$ and $\aabz(\boldsymbol{\theta}) \in \complexset{\Mz}{1}$, with their elements given by $[\aabx(\boldsymbol{\theta})]_n= e^{\jmath \frac{2 \pi}{\lambda}   \dx n \cos(\thetael) \sin(\thetaaz)} $ and $[\aabz(\boldsymbol{\theta})]_n = e^{\jmath \frac{2 \pi}{\lambda}   \dz n \sin(\thetael)}$, respectively. Here, $\dx$ and $\dz$ represent the spacing between array elements. The complete spatial-domain steering vector is obtained through the Kronecker product: $\aab(\boldsymbol{\theta}) = \aabx(\boldsymbol{\theta}) \otimes \aabz(\boldsymbol{\theta})$. Additionally, the frequency-domain steering vector $\dd(\tau) \in \complexset{S}{1}$ is defined element-wise as $[\dd(\tau)]_{\kappa}=e^{-\jmath2\pi \kappa\Delta{f}\tau}$, while $\nnm_k$ accounts for the noise component in the tensor representation.

\subsubsection{Tensor Decomposition With Spatial Augmentation}
To extract the path parameters from the tensor expression in \eqref{eq_yy_3d}, we utilize a high-resolution channel estimation method that is based on tensor decomposition, as proposed in \cite{ge2024v2x}. The key idea is to overcome the rank deficiency issue inherent in conventional CP decomposition (CPD) approaches. To achieve this, we reshape the original observation tensor into a new form, $\yymt_k \in \complexsett{\Mz(n_z+1)}{\Mx(n_x+1)}{V}$, where $V = S - n_z - n_x$. This transformation is achieved by augmenting the spatial dimensions using frequency-domain data, following the spatial augmentation (SA) framework in \cite[Eq.~(15)]{ge2024v2x}. After applying SA, the signal tensor can be expressed as 
\begin{align} \label{eq_yymt_3d}
     &\widetilde{\yym}_k = \sum _{t=1}^{T_{k}}\rho^{t}_{k} \aabzw(\boldsymbol{\theta}_{k}^{t},\tau_k^{t}) \circ \aabxw(\boldsymbol{\theta}_{k}^{t},\tau_k^{t}) \circ \widetilde{\dd}(\tau_k^{t}) + \widetilde{\nnm}_k \,,
 \end{align}
 where $\widetilde{\nnm}_k$ denotes the reshaped form of $\nnm_k$, and
 \begin{subequations} \label{eq_sa_steering}
 \begin{align}
     \aabzw(\boldsymbol{\theta},\tau) &= \aabz(\boldsymbol{\theta}) \otimes  [\dd(\tau)]_{1:n_z+1} \in \complexset{\Mz(n_z+1)}{1} ~, \\
     \aabxw(\boldsymbol{\theta},\tau) &= \aabx(\boldsymbol{\theta}) \otimes  [\dd(\tau)]_{1:n_z+1} \in \complexset{\Mx(n_x+1)}{1} ~, \\
     \widetilde{\dd}(\tau) &= [\dd(\tau)]_{1:V}  \in \complexset{V}{1}~.
 \end{align}
 \end{subequations} 
In this formulation, the augmented spatial-frequency domain representations $\aabzw$ and $\aabxw$ allow for improved identifiability of the path parameters by effectively increasing the rank of the observed tensor, thereby enabling more accurate decomposition and estimation.
Following this, we proceed by applying CPD to the augmented tensor $\widetilde{\yym}_k$ as expressed in \eqref{eq_yymt_3d}. The decomposition problem is formulated as
\begin{subequations} \label{eq_cp_sa}
\begin{align}
    \min_{ \{\aabzi, \aabxi, \ddi \}_{t=1}^{T_{k}} } & ~\normsmall{\hat{\yym} - \widetilde{\yym}_k}_F^2
    \\
    \mathrm{s.t.}&~~  \hat{\yym} = \sum _{t=1}^{T_{k}} \aabzi \circ \aabxi \circ \ddi ~.
\end{align}
\end{subequations}  
Using the structure of the augmented steering vectors introduced in \eqref{eq_sa_steering}, the path-specific parameters for all components at time step $k$ can be retrieved from the decomposition results in \eqref{eq_cp_sa} as follows:
{\allowdisplaybreaks
\begin{subequations}\label{eq_cpd_sa_est}
\begin{align} \label{eq_tauhatl_cpd_sa}
 \hat{\tau}_k^{t} &=  -\frac{
    \angle \big( [\JJ_{1} \ddi]\herm [\JJ_{2} \ddi]  \big)    }{ 2 \pi \deltaf} ~, \\ \hat{\theta}^{t}_{\text{el},k}&= \arg \max_{\thetael} \abs{ (\aabzi)\herm  \aabzw(\boldsymbol{\theta}, \hat{\tau}_k^{t})   }^2 ~,    \\
    \hat{\theta}^{t}_{\text{az},k} &= \arg \max_{\thetaaz} \abs{ (\aabxi)\herm  \aabxw(\thetaaz, \hat{\theta}^{t}_{\text{el},k}, \hat{\tau}_k^{t})   }^2 ~.
\end{align}
\end{subequations}}Here, $\JJ_{1}$ and $\JJ_{2}$ refer to selection matrices that extract the first and final $V - 1$ entries of any right-multiplied vector, respectively. Once the delay and angular parameters have been estimated, the path gains are inferred via a least-squares approach, i.e., $\hat{\boldsymbol{\rho}}_{k} = \AAb_k^{\dagger} \breve{\boldsymbol{y}}_k$. In this expression, $\AAb_k \in \complexset{\Mz \Mx S}{T_k}$ is a matrix whose $t$-th column is defined as
\begin{align}
 [\AAb_k]_{:,t} = \dd(\hat{\tau}_k^{t}) \otimes \aabx(\hat{\theta}^{t}_{\text{az},k}, \hat{\theta}^{t}_{\text{el},k}) \otimes \aabz(\hat{\theta}^{t}_{\text{az},k},\hat{\theta}^{t}_{\text{el},k})    \,,
\end{align}
$\breve{\boldsymbol{y}}_k = \vect\left(\yym_k\right) \in \complexset{\Mz \Mx S}{1}$, $\hat{\boldsymbol{\rho}}_{k} \in \complexset{T_k}{1} $ with $[\hat{\boldsymbol{\rho}}_{k}]_t = \hat{\rho}_{k}^{t}$ denoting the estimate of the path gain $\rho^{t}_{k}$ in \eqref{eq_yy_3d}, and $(\cdot)^{\dagger}$ denotes Moore-Penrose pseudo-inverse.

\subsection{User Positioning and Landmark Mapping} \label{positioning and mapping}
\subsubsection{UE Positioning} 
The ESPRIT channel parameter estimator outputs a set of estimated paths for each UE-BS pair, denoted by $\{\hat{\tau}_{k}^{t},\hat{\boldsymbol{\theta}}_{k}^{t}\}_{t\in\{1,\dots,\hat{T_{k}}\}}$, where $\hat{T}_{k}$ is the total number of estimated paths. These paths may include both \ac{LoS} and/or \ac{NLoS} components. For UE positioning, only the \ac{LoS} paths are used, since the \acp{IP} of the \ac{NLoS} paths are unknown. We further assume that the UEs move on a flat ground plane in the considered scenarios, their heights are known, and thus only the horizontal coordinates, $x_{\text{UE},k}$ and $y_{\text{UE},k}$, need to be estimated. Given the measurement $\boldsymbol{z}_{k}=[\tau_{k,\text{LoS}}^{t},\boldsymbol{\theta}_{k,\text{LoS}}^{t}]$ for the \ac{LoS} path, which is assumed to be the shortest estimated path, and its associated covariance matrix $\boldsymbol{R}_{k}$, the unknown UE coordinates $(x_{\text{UE},k}, y_{\text{UE},k})$ can be estimated using a \ac{ML} estimator, defined as
\begin{align} \label{eq_ml}
    &\hat{x}_{\text{UE},k},\hat{y}_{\text{UE},k} = \\ &
    \arg \min_{x_{\text{UE},k},y_{\text{UE},k}} \,  (\boldsymbol{z}_{k} - \boldsymbol{h}(\boldsymbol{x}_{\text{UE},k},\boldsymbol{x}_{\text{BS}}))^{\mathsf{T}} \boldsymbol{R}_{k}^{-1} (\boldsymbol{z}_{k} - \boldsymbol{h}(\boldsymbol{x}_{\text{UE},k},\boldsymbol{x}_{\text{BS}})),\nonumber
\end{align}
where $\boldsymbol{h}(\boldsymbol{x}_{\text{UE},k},\boldsymbol{x}_{\text{BS}})$ denotes the measurement function that maps the positions of the UE and the BS to the TOA and AOA of the LoS path. The estimation problem can be solved using gradient descent, initialized with a closed-form \ac{LS} solution as described in \cite{zhu2009simple}.

\subsubsection{Landmark Mapping} 
After estimating the UE position, we proceed to map the environment by estimating the \acp{IP} associated with the remaining \ac{NLoS} paths. Since each resolved path provides only three observable parameters, we can estimate at most one unknown \ac{IP}. Determining multiple \acp{IP} for multi-bounce paths is infeasible without additional information, making the problem underdetermined. 
In addition, it is not known whether each \ac{NLoS} path corresponds to a single-bounce or a multi-bounce path. Therefore, we conservatively treat all remaining \ac{NLoS} paths as single-bounce paths and estimate the corresponding \ac{IP} for each path accordingly, even though some, in fact, are multi-bounce paths. 

For a \ac{NLoS} path with measurements $\{\tau_k^{t},\boldsymbol{\theta}_{k}^{t}\}$, the signal is modeled as propagating in the order $\textrm{UE} \rightarrow \textrm{IP} \rightarrow \textrm{BS}$, where $\boldsymbol{x}_{\text{IP},k}^{t}$ denotes the position of the corresponding \ac{IP}. The \ac{IP} lies along the line originating from the BS in the direction of $\boldsymbol{\theta}_{k}^{t}$, and the total path length from the UE to the BS via the IP is constrained by the measured delay $\tau_k^{t}$. Based on these geometric constraints, the position $\boldsymbol{x}_{\text{IP},k}^{t}$ can be determined by
\begin{align}
\begin{cases}
   &\lVert\boldsymbol{x}_{\text{IP},k}^{t}-\boldsymbol{x}_{\text{UE},k}^{t}\rVert + \lVert\boldsymbol{x}_{\text{IP},k}^{t}-\boldsymbol{x}_{\text{BS}}\rVert = \tau_k^{t}, \\&(\boldsymbol{x}_{\text{IP},k}^{t}-\boldsymbol{x}_{\text{BS}})^{\top}\boldsymbol{u}^{t}_{k}= \lVert\boldsymbol{x}_{\text{IP},k}^{t}-\boldsymbol{x}_{\text{BS}}\rVert, 
\end{cases}
\label{eq:so_interaction_point}
\end{align}
where  $\boldsymbol{u}^{t}_{k}$  denotes the unit direction vector determined by $\boldsymbol{\theta}_{k}^{t}$
\begin{align}
    \boldsymbol{u}^{t}_{k} &= \boldsymbol{R}(\boldsymbol{\psi}_\text{BS})^{\top} \begin{bmatrix}
        \cos(\theta_{\text{az},k}^{t})\cos(\theta_{\text{el},k}^{t}) \\\sin(\theta_{\text{az},k}^{t})\cos(\theta_{\text{el},k}^{t}) \\ \sin(\theta_{\text{el},k}^{t})
    \end{bmatrix}, \label{eq:ue_unit_vector} 
\end{align}
 with $\boldsymbol{R}(\boldsymbol{\psi}_\text{BS})$ denoting the rotation matrix that transforms coordinates from the global reference frame to the local coordinate system of the BS \ac{URA}. The first entry in \eqref{eq:so_interaction_point} ensures that $\boldsymbol{x}_{\text{IP},k}^{t}$ lies on an ellipse with foci at $\boldsymbol{x}_{\text{BS}}$ and $\boldsymbol{x}_{\text{UE},k}$. The second entry specifies that the IP lies on a half-line originating from $\boldsymbol{x}_{\text{BS}}$ in the direction of $\boldsymbol{u}^{t}_{k}$. Consequently, $\boldsymbol{x}_{\text{IP},k}^{t}$ is determined as the intersection point of this half-line and the ellipse. An illustrative representation of this geometric interpretation is shown in Fig.~\ref{fig:ellipse}. While a closed-form solution to \eqref{eq:so_interaction_point} exists, the derivation is omitted for brevity. This process is repeated for all resolved \ac{NLoS} paths to estimate the corresponding \acp{IP}.

\section{Post-processing} \label{Sec:postprocessing}
At each time step, each BS sends a set of estimated \acp{IP} to a central fusion center, forming a global set $\mathcal{D}$. This section introduces a three-phase method to estimate extended landmarks and detect outliers—i.e., estimated \acp{IP} of multi-bounce paths mistakenly treated as single-bounce reflections.

\subsection{Phase I: DBSCAN for Clustering}
An \ac{IP} represents the point where a signal hits a landmark. As such, \acp{IP} corresponding to paths bounced from the same landmark should lie on that landmark. This allows us to estimate a landmark that best fits the set of \acp{IP} originating from the same source. However, the origin source of each \ac{NLoS} path is initially unknown, making it infeasible to directly estimate landmarks from the complete set of collected paths. To address this, we first cluster all \acp{IP} that likely originate from the same source, based on the assumption that such points are spatially close to one another. For this clustering, we employ the \ac{DBSCAN} algorithm \cite{DBSCAN96}, which groups points based on density-reachability and classifies them either as members of a cluster or as outliers.

The \ac{DBSCAN} algorithm is controlled by two parameters $\epsilon$ and $N_{\min}$. A core point is defined as a point with at least $N_{\min}$ points (including itself) within the distance of $\epsilon$, where we use the Euclidean distance in this paper. A point is directly reachable from a core point if the distance between two points is within $\epsilon$, and points are only directly reachable from core points. Moreover, a point is reachable from a given core point if it can be reached through a series of core points starting from the given core point, where the current core point is directly reachable from the last core point. Please note that all the points on the path from the given core point to the target point  should be core points, with the possible exception of the target point. Any points that are not reachable from any other point are defined as outlier points. The \ac{DBSCAN} algorithm clusters a core point with all its reachable points, no matter core or non-core points, to create a cluster, and forms different clusters and picks up all outlier points. If we input $\mathcal{D}$, the output of the DBSCAN algorithm will be a list of point clusters $\mathcal{L}_{1}, \dots,\mathcal{L}_C$, and set of outlier points $\mathcal{P}$.

\subsection{Phase II: Convex Hull Extraction}
Applying DBSCAN to $\mathcal{D}$ yields clusters $\mathcal{L}_{c} = \{\boldsymbol{x}_{\text{IP}}^{1}, \ldots, \boldsymbol{x}_{\text{IP}}^{|\mathcal{L}_{c}|}\} \subset \mathcal{D}$, where each cluster represents a group of estimated \acp{IP} that are spatially close and likely to originate from the same physical landmark. To extract a geometric boundary for subsequent tasks, such as occlusion detection or shape analysis, we compute the convex hull of each cluster. The convex hull, denoted as $\text{CH}(\mathcal{L}_{c})$ for the defined cluster $\mathcal{L}_{c}$, is the smallest convex set that encloses all points in $\mathcal{L}_{c}$, defined as
\begin{equation}
\label{eq:convhull}
\text{CH}(\mathcal{L}_{c}) 
= \left\{ \boldsymbol{x} \in \mathbb{R}^3 \,\middle|\,
\boldsymbol{x} = \sum_{n=1}^{|\mathcal{L}_{c}|} \alpha_n \boldsymbol{x}_{\text{IP}}^{n},\;
\alpha_n \geq 0,\;
\sum_{n=1}^{|\mathcal{L}_{c}|} \alpha_n = 1 
\right\}.
\end{equation}
Any point $\boldsymbol{x}$ within the convex hull is represented as a convex combination of the points in the cluster. Efficient algorithms such as Quickhull \cite{barber1996quickhull} are commonly used to compute $\text{CH}(\mathcal{L}_{c})$, yielding a compact and computationally efficient boundary representation for each cluster, even in the presence of internal irregularities. While the convex hull may not capture concavities in the original shape, its simplicity and efficiency make it well-suited for many real-time 3D applications. Alternative approaches, such as concave hulls, can provide more accurate representations of complex geometries, but their exploration lies outside the scope of this paper.

\subsection{Phase III: Outlier Removal}
After extraction, a convex hull can be recovered for each cluster, capturing geometric properties of the associated landmarks, such as location, size, and shape. However, not all clusters produced by DBSCAN necessarily correspond to real physical landmarks. Many multi-bounce paths may be misinterpreted as single-bounce reflections, resulting in the formation of false landmarks. Distinguishing between single- and multi-bounce paths based solely on the measurements—angles, delays, or the 3D positions of the estimated \acp{IP}—is inherently challenging, as this information does not directly reveal the nature of the underlying bounces.

A key geometric cue for identifying false landmarks is occlusion: IPs generated by multi-bounce paths often appear \textit{behind} real landmarks due to longer propagation distances. Although such paths may yield similar AOA  compared to single-bounce paths, they typically exhibit greater TOA, placing the estimated IPs farther from both the BS and UE. These false IPs are often occluded by real landmarks. To eliminate such false landmarks, we introduce a visibility-based test for each estimated IP $\boldsymbol{x}_{\text{IP}}^{n}$. Let $\boldsymbol{x}_{\text{BS}}$ and $\boldsymbol{x}_{\text{UE},k}$ denote the corresponding BS and UE positions, respectively, and we define the BS-IP and UE-IP line segments as
\begin{align}
    \mathcal{S}_{\text{BS}}^{n} &= \left\{ \boldsymbol{x}(\alpha) = \boldsymbol{x}_{\text{BS}} + \alpha\left(\boldsymbol{x}_{\text{IP}}^{n} - \boldsymbol{x}_{\text{BS}}\right) \mid 0 < \alpha < 1 \right\}, \\
    \mathcal{S}_{\text{UE}}^{n} &= \left\{ \boldsymbol{x}(\alpha) = \boldsymbol{x}_{\text{UE}} + \alpha\left(\boldsymbol{x}_{\text{IP}}^{n} - \boldsymbol{x}_{\text{UE},k}\right) \mid 0 < \alpha < 1 \right\}.
\end{align}
We check whether $\mathcal{S}_{\text{BS}}^{n}$ or $\mathcal{S}_{\text{UE}^{n}}$ intersects with any other convex hull $\mathrm{CH}(\mathcal{L}_c)$ formed from another IP cluster $\mathcal{L}_c$. An IP is retained only if both lines are unobstructed:
\begin{equation}
    \mathcal{S}^{n}_{\text{BS}} \cap \mathrm{CH}(\mathcal{L}_c) = \emptyset \text{ and } \mathcal{S}^{n}_{\text{UE}} \cap \mathrm{CH}(\mathcal{L}_c) = \emptyset, \quad \forall \, \mathcal{L}_c, \boldsymbol{x}_{\text{IP}}^{n}\notin\mathcal{L}_c.
\end{equation}
The remaining visible IPs are used to construct convex hulls that represent the estimated landmarks.

\section{Results}

\subsection{Simulation Environment}
\begin{figure}
    \centering
    \includegraphics[width=0.8\linewidth]{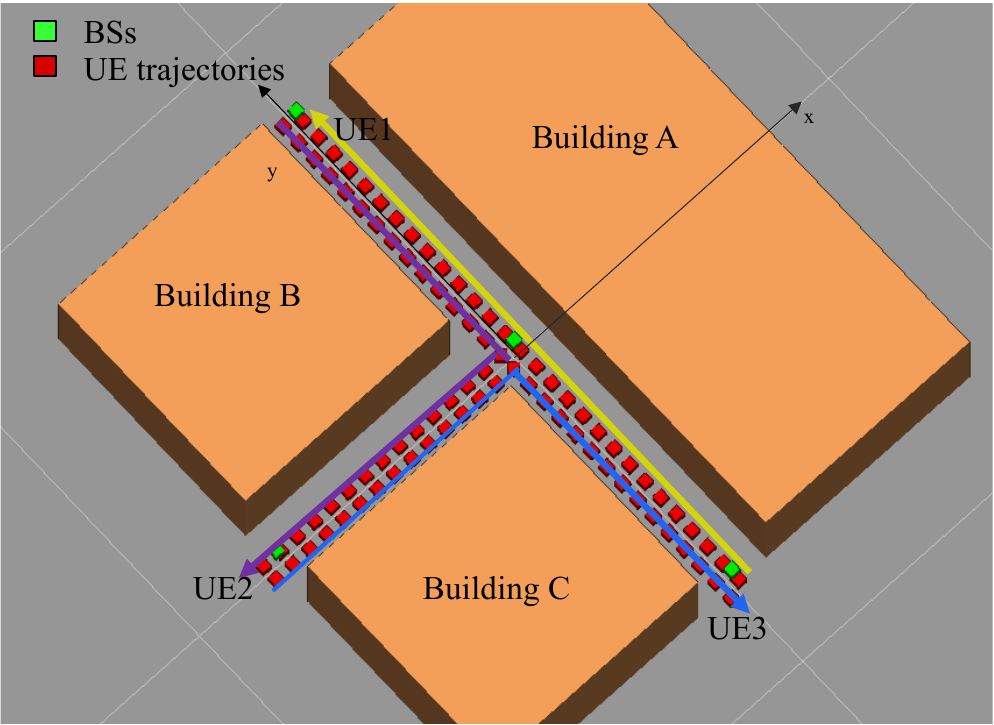}
    \vspace{-2mm}
    \caption{The ray-tracing simulation is conducted in a T-junction environment, featuring four fixed \acp{BS} and three UEs, each following its own trajectory over time. Building A is centered at $[40 \, \text{m},0 \, \text{m},15\, \text{m}]^{\top}$ and measures $140 \, \text{m}$ in length, $50 \, \text{m}$ in width, and $30 \, \text{m}$ in height. Buildings B and C are centered at $[40 \, \text{m},40 \, \text{m},15\, \text{m}]^{\top}$ and $[40 \, \text{m},-40 \, \text{m},15\, \text{m}]^{\top}$, respectively, each measuring $60 \, \text{m}$ in length, $60 \, \text{m}$ in width, and $30 \, \text{m}$ in height.}
    \label{fig:raytracing_env}\vspace{-6mm}
\end{figure}

We consider an urban vehicular scenario situated at a T-junction, as illustrated in Fig.~\ref{fig:raytracing_env}. The environment consists of three surrounding buildings with brick surfaces. Four BSs and three vehicles that act as UEs are deployed. The BSs are located at $[0 \, \text{m},0 \, \text{m},15\, \text{m}]^{\top}$, $[0 \, \text{m}, -70 \, \text{m},15\, \text{m}]^{\top}$, $[0 \, \text{m}, 70 \, \text{m},15\, \text{m}]^{\top}$, and $[-70 \, \text{m}, 0 \, \text{m},15\, \text{m}]^{\top}$, respectively. The UEs follow distinct trajectories over time: UE1 moves from $[2 \, \text{m},-70 \, \text{m}]^{\mathsf{T}}$ to $[2 \, \text{m},70 \, \text{m}]^{\mathsf{T}}$. UE2 starts at $[-70 \, \text{m},-2 \, \text{m}]^{\mathsf{T}}$, turns right at $[-2 \, \text{m},-2 \, \text{m}]^{\mathsf{T}}$, and ends at $[-2 \, \text{m},-70 \, \text{m}]^{\mathsf{T}}$. UE3 starts at $[-2 \, \text{m},70 \, \text{m}]^{\mathsf{T}}$, turns right at $[-2 \, \text{m},2 \, \text{m}]^{\mathsf{T}}$, and ends at $[-70 \, \text{m},2 \, \text{m}]^{\mathsf{T}}$. Along each trajectory, samples are taken at 5~m intervals. Each BS is equipped with an $8 \times 8$ multi-panel antenna array, while each UE has a single antenna mounted at a fixed height of $1.5\, \text{m}$, which is assumed to be known. The antenna radiation patterns follow the specifications in \cite{ge2024v2x}. The coverage radius of each BS is set to $70 \, \text{m}$. A UE communicates only with the BSs within the coverage ranges of BSs, ensuring LoS availability for all active BS-UE pairs.

The carrier frequency is set to $27.2~\text{GHz}$. The system employs OFDM pilot signals composed of 12 symbols, each with constant amplitude. The total active bandwidth is $400~\text{MHz}$, with a subcarrier spacing of $120~\text{kHz}$. The transmit power is configured to $10~\text{dBm}$, while the noise spectral density is set to $-174~\text{dBm/Hz}$ and the receiver noise figure to $8~\text{dB}$. Ground-truth channel data is obtained using the REMCOM Wireless InSite\textregistered ray-tracing simulator \cite{WirelessInSite}. At each UE location and time step, 50 propagation paths are retained per BS. On average, each snapshot contains approximately 9 single-bounce paths and 40 double-bounce paths per BS. For clustering using the DBSCAN algorithm, $\epsilon$ is set to 3~m, and $N_{\min}$ is set to 5. 100 Monte Carlo (MC) simulations are performed.

Each simulation starts with generating propagation paths using ray-tracing data, followed by channel parameter estimation described in Section~\ref{channel_estimator}. The positioning and mapping method in Section~\ref{positioning and mapping} is then applied to estimate UE positions and the corresponding IPs. Post-processing in Section~\ref{Sec:postprocessing} clusters IPs, reconstructs convex hulls, and removes outliers. Positioning performance is evaluated using the sub-meter accuracy rate and the \ac{MAE} across all UEs over time and 100 MC simulations. Mapping performance is assessed qualitatively through visual comparison with the ground-truth map, and quantitatively by measuring the outlier removal rate.

\subsection{Results and Discussion}
We begin by evaluating the positioning performance of all UEs, using the channel parameter estimates corresponding to the \ac{LoS} path. The \ac{LoS} path is taken to be the shortest among the resolved paths provided by the channel parameter estimator at each time step.
The results indicate that sub-meter accuracy is achieved in 95.9\% of the cases, with an overall MAE of 0.21~m. This high positioning accuracy is primarily attributed to the wide signal bandwidth and large antenna array, which offer fine angular and delay resolution, as well as the use of a high-precision channel parameter estimator.

\begin{figure}
\center
\input{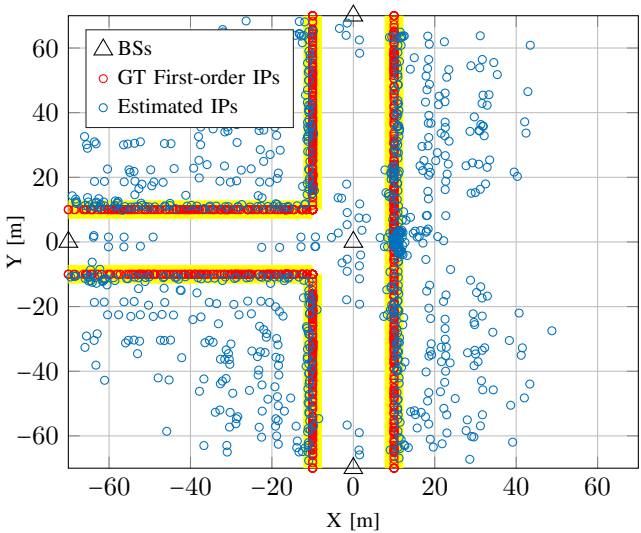}
\vspace{-3mm} 
\caption{2D visualization of the mapping results, showing the estimated IPs corresponding to all resolved channel paths for all UE-BS pairs. The ground truth (denoted as ``GT" in the figure) consists of the true IPs from all single-bounce paths across all UE-BS pairs and time steps. All ground-truth IPs lie on the building facades, which are indicated by yellow lines. The results correspond to a single MC simulation.}
\label{WOPOST}
\vspace{-3mm} 
\end{figure}

Next, we evaluate the mapping performance and demonstrate how the proposed post-processing methods enhance landmark reconstruction and outlier removal. Fig.~\ref{WOPOST} illustrates the mapping results \textit{without} applying our post-processing scheme, where the estimated IPs for all resolved paths are shown. All first-order IPs corresponding to single-bounce paths, and ground-truth map are also shown. From the figure, we observe that some estimated IPs align well with the first-order IPs and ground truth map, indicating that certain estimated paths do correspond to true single-bounce paths. In these cases, our mapping method performs effectively, producing IPs that are spatially close to the true IPs. Specifically, out of 870 estimated IPs, 379 lie within 2~m of the ground-truth facades. However, we also notice a significant number of estimated IPs located far from the real first-order IPs and building facades. These errors arise from paths that are either clutter-induced or involve multiple bounces. Since these paths do not conform to the single-bounce assumption, their corresponding IPs are misleading and should be excluded from the mapping process.

\begin{figure}
\center
\input{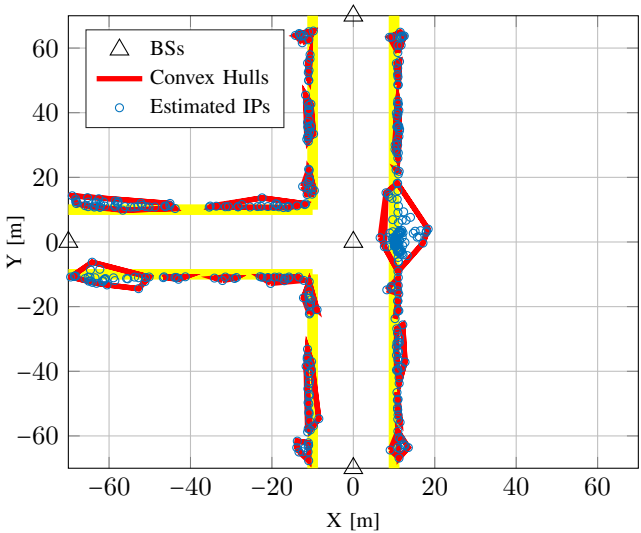}
\vspace{-3mm} 
\caption{2D visualization of the mapping results after applying the proposed post-processing scheme. The results correspond to a single MC simulation.}
\label{WPOST}
\vspace{-5mm} 
\end{figure}

Fig.~\ref{WPOST} shows the mapping performance with the proposed post-processing scheme. It is evident that most of the spurious IPs have been successfully removed. In total, 43.1\% of the estimated IPs are discarded, and the remaining IPs align well with the building facades, demonstrating the effectiveness of our outlier removal method. Furthermore, Fig.~\ref{WPOST} also shows the reconstructed convex hulls, which closely match the actual building facades. These convex hulls provide richer geometric information beyond individual IPs, enabling estimation of the approximate size and shape of the facades. Some discontinuities are observed compared to the ground-truth, primarily due to missing or sparse IP estimates in those regions. In such cases, DBSCAN may classify the sparse IPs as outliers. Additionally, larger channel parameter estimation errors occur when the azimuth angle is near zero, leading to increased UE positioning and IP mapping errors, and resulting in a distorted cluster, for example around $[0 \, \text{m},10 \, \text{m}]^{\top}$. The mapping performance is expected to improve further with more UEs and/or BSs in the environment, as additional radio paths offer more opportunities for signal interaction with the environment and thus provide more spatial information.

\vspace{-2mm}

\section{Conclusions}
In this paper, we address the limitations of existing environmental modeling approaches in \ac{ISAC}, which typically focus on isolated IPs. Our goal is to enable rich, radio-based environmental mapping for future 6G networks by moving beyond simplistic point-based representations and leveraging more expressive features to enhance sensing performance. To achieve this, we propose an end-to-end cooperative uplink framework involving multiple BSs and UEs. Within this framework, UEs are localized, extended landmarks are estimated, and a visibility-based outlier removal method is introduced to eliminate clutter caused by multi-bounce paths. Simulation validation using realistic ray-tracing data demonstrates the effectiveness of the proposed method for high-precision positioning and environmental reconstruction in complex wireless scenarios.

\scriptsize{
\section*{Acknowledgment}
This work has been supported by the SNS JU project 6G-DISAC under the EU's Horizon Europe research and innovation Program under Grant Agreement No.~101139130, SNS JU project 6G-MUSICAL under Grant Agreement No.~101139176 and the Swedish Research Council (VR) through the project 6G-PERCEF under Grant 2024-04390. The authors wish to thank Remcom for providing Wireless InSite\textregistered  ray-tracer.
}
\bibliography{IEEEabrv,Bibliography}

\end{document}